\documentclass[preprint,showpacs,preprintnumbers,endfloats*,amsmath,amssymb,linenumbers,prb]{revtex4-1}
\usepackage{graphicx}
\usepackage{dcolumn}
\usepackage{bm}
\newcommand{\vecvar}[1]{\mbox{\boldmath$#1$}}

\begin{document}

\preprint{PRESAT-8601}

\title{Fully spin-dependent transport of triangular graphene flakes}

\author{Tomoya Ono$^1$, Tadashi Ota$^1$, and Yoshiyuki Egami$^2$}
\affiliation{$^1$Department of Precision Science and Technology, Osaka University, Suita, Osaka 565-0871, Japan\\
$^2$Nagasaki University Advanced Computing Center, Nagasaki University, Bunkyo-machi, Nagasaki 852-8521, Japan }

\date{\today}

\begin{abstract}
The magnetic moment and spin-polarized electron transport properties of triangular graphene flakes surrounded by boron nitride sheets (BNC structures) are studied by first-principles calculation based on density functional theory. Their dependence on the BNC structure is discussed, revealing that small graphene flakes surrounded by large BN segments have a large magnetic moment. When the BNC structure is suspended between graphene electrodes, the spin-polarized charge density distribution accumulates at the edge of the graphene flakes and no spin polarization is observed in the graphene electrodes. We also found that the BNC structure exhibits perfectly spin-polarized transport properties in a wide energy window around the Fermi level. Our first-principles results indicate that the BNC structure provides for the new possibilities for the electrical control of spin.
\end{abstract}

\maketitle
\section{Introduction}
\label{sec:intro}
Graphene,\cite{graphene} a two-dimensional monolayer honeycomb structure of carbon, is known to exhibit a rich variety of electronic structures and is one of the most promising new materials for future nanoelectronics. The electronic reconstruction of graphenes induced at boundaries can give rise to metal or insulator states,\cite{hamada} magnetism,\cite{zgnr} or even superconductivity.\cite{heersche} The discovery of zigzag graphene nanoribbons,\cite{zgnr,nakada} in which an opposite spin orientation crosses the ribbon between ferromagnetically ordered edge states on each edge, through theoretical calculations has attracted a great deal of interest\cite{tombros,karpan} in spintronics applications based on graphene-based materials. The possibility of controlling electron transport by means of the spin degree of freedom has recently attracted attention, because spintronics devices may have potential for applications in future commercial electronics and for generating insights into the fundamental properties of electron spin in solids. From the viewpoint of the development of highly efficient spintronics devices, the spin-filter effect, which can be used for the efficient injection of spins into magnetic junctions, is an important issue of concern and debate. Okada and Oshiyama\cite{okada} studied the spin polarizations of two-dimensional structures composed of boron, nitrogen, and carbon, in which triangular graphene flakes are surrounded by boron nitride (BN) sheets (referred to as BNC structures hereafter), through first-principles calculations and found that flat-band states can be observed around the Fermi level and the BNC structures are ferromagnetically polarized. Zheng {\it et al.}\cite{antidot} examined the spin transport properties of graphene antidots, i.e., graphenes with rectangular or triangular holes, and observed the spin polarization of electron current in triangular antidots. However, the energy window where current is perfectly spin-polarized is rather small in their system because the energy of the edge states contributing to spin-polarized current does not shift significantly around the antidots.

We study the relationship between the magnetic moment of BNC structures and their sizes. The spin-dependent transport properties of graphene/BNC/graphene (G/BNC/G) structures, where BNC structures are sandwiched with graphene electrodes, are also examined. We found that small graphene flakes surrounded by long BN segments exhibit a large magnetic moment and that the BNC structures with small flakes exhibit a fully spin-polarized current with a large energy range of the incident electrons.

All calculations are done within the framework of density functional theory (DFT)\cite{dft} using a real-space finite-difference approach,\cite{chelikowsky,book} which makes it possible to carry out calculations with a high degree of accuracy by combining them with a timesaving double-grid technique. \cite{book,tsdg} Valence electron-ion interaction is described using norm-conserving pseudopotentials\cite{norm} generated by the scheme proposed by Troullier and Martins.\cite{tm} Exchange and correlation effects are treated within the local spin density approximation\cite{lda} of DFT.

\section{Results and discussion}
\label{sec:rd}
\subsection{Magnetic moment of graphene flakes}
\begin{figure*}
\includegraphics{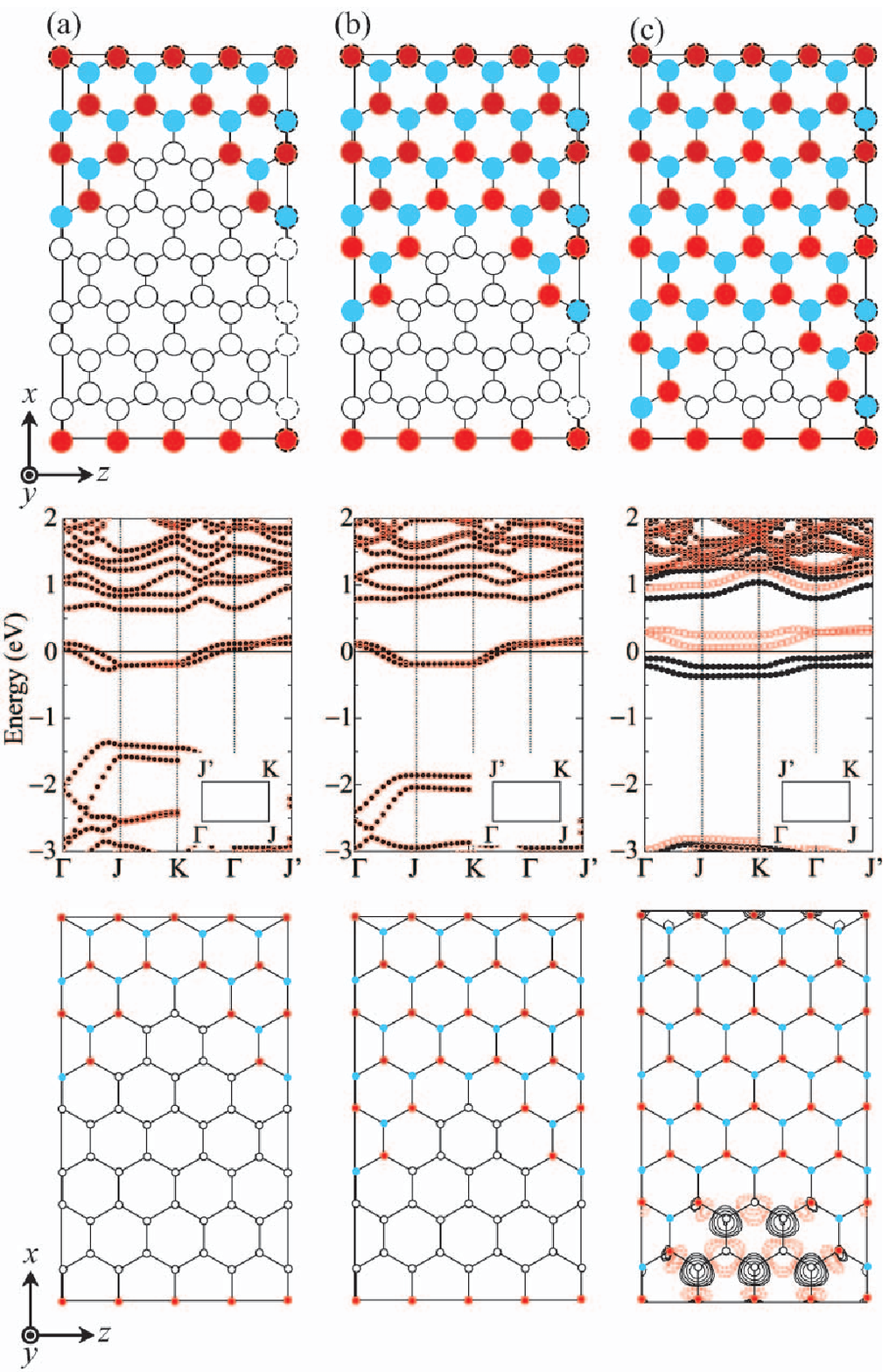}
\caption{(color online) Top view of computational models (top panel), electronic band structures (middle panel), and contour plots showing difference between up-spin and down-spin charge density distributions (bottom panel). (a) represents a large graphene flake model, (b) represents a medium graphene flake model, and (c) represents a small graphene flake model. White, light blue (gray), and red (black) circles correspond to C, B, and N atoms. The atoms enclosed by dashed circles are in the neighboring supercell. In the band structures, closed and open circles are up-spin and down-spin bands, respectively. In the contour plots, positive values of spin density are indicated by solid lines and negative values by dashed lines. Each contour represents twice or half the density of the adjacent contour lines. The lowest contour represents 3.29 $\times$ $10^{-3}$ $e$/\AA$^3$.}
\label{fig:1}
\end{figure*}

\begin{figure*}
\includegraphics{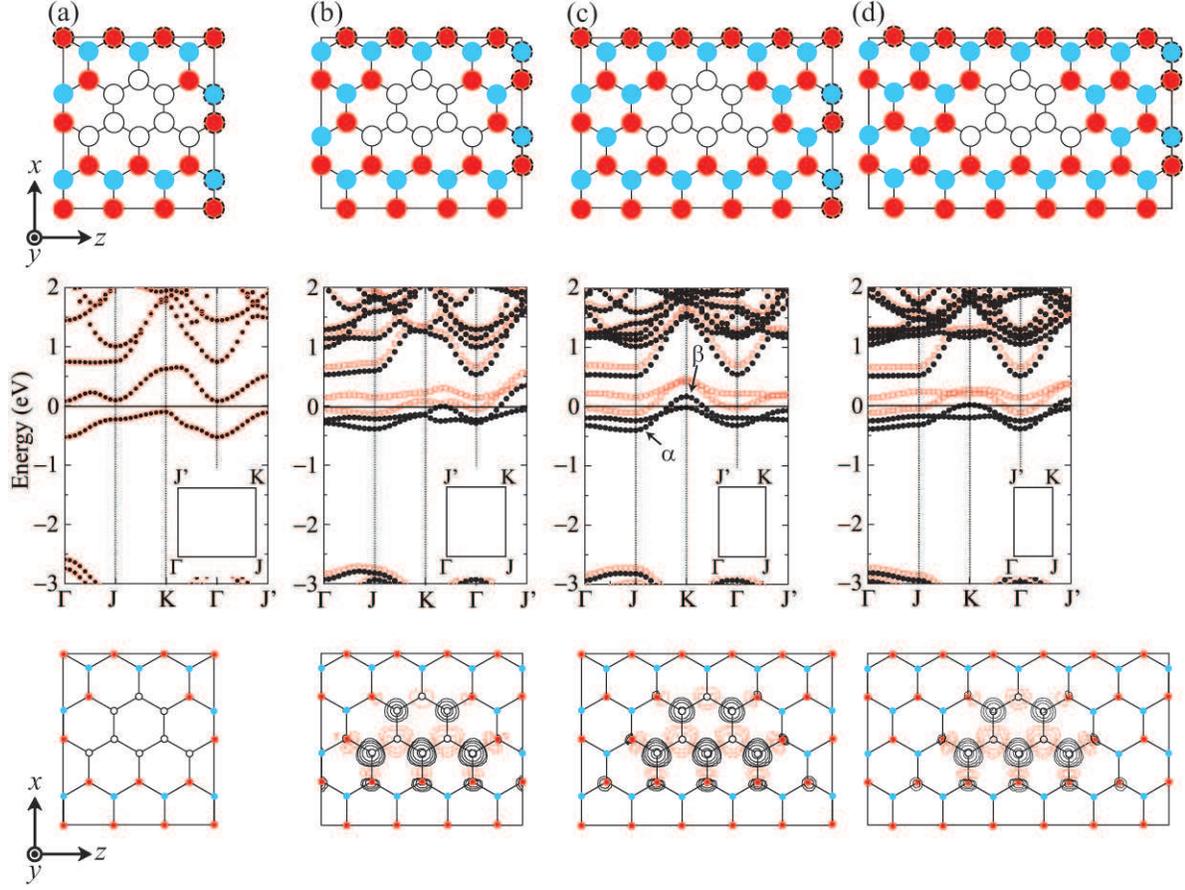}
\caption{(color online) Top view of computational models (top panel), electronic band structures (middle panel), and contour plots showing difference between up-spin and down-spin charge density distributions (bottom panel). (a), (b), (c), and (d) correspond to short, medium, long, and very long distances between graphene flakes. The symbols have the same meanings as those in Fig.~\ref{fig:1}.}
\label{fig:2}
\end{figure*}

Let us first consider three periodic BNC structures to investigate the effect of the size of carbon regions on the magnetic moment. Figures~\ref{fig:1}(a), \ref{fig:1}(b), and \ref{fig:1}(c) show the computational models we employed here. Here and hereafter, we refer to the BNC structures in Figs.~\ref{fig:1}(a), \ref{fig:1}(b), and \ref{fig:1}(c) as models 1(a), 1(b), and 1(c), respectively. There are $64$ atoms in the supercell. Periodic boundary conditions are imposed on all directions, and a repeating sheet model is separated by 9.0 \AA \hspace{1mm} in each layer for all calculations presented in this section. The Brillouin zone is sampled using a $2 \times 1 \times 8$ $k$-point grid. The lattice constant of graphene is set at 1.41 \AA \hspace{1mm} with a real-space grid spacing of $\sim$ 0.18 \AA, and structural optimization is performed until the remaining forces are less than 5 mRy/\AA. The calculated magnetic moments of the BNC structures are listed in Table~\ref{tbl:1}. In addition, the electronic band structures are plotted in the middle panels of Fig.~ \ref{fig:1}. The magnetic moment of the BNC structures with the large graphene flakes [models 1(a) and 1(b)] are zero, and the moment becomes $S_0$=2, as expected from Lieb's theorem for a biparticle lattice,\cite{lieb} when the graphene segment becomes small. We then examine variations in the magnetic moments as a function of the distance in the $z$ direction between graphene flakes in the neighboring supercell. Figures~\ref{fig:2}(a), \ref{fig:2}(b), \ref{fig:2}(c), and \ref{fig:2}(d) show the calculated atomic configurations. Integration over the Brillouin zone is carried out using a $4 \times 1 \times 8$ $k$-point grid. The calculated magnetic moments and electronic band structures are shown in Table~\ref{tbl:1} and in the middle panels of Fig.~\ref{fig:2}, respectively. No spin polarization is observed in the BNC structure with the shortest graphene-flake distance [model 2(a)]. The magnetic moment increases with increasing in the graphene-flake distance because the BN regions insulate the triangular flakes so that the behavior of the edge states is flat. The moment $S_0$ approaches a noninteger value of $\sim 1.5$ and does not become 2.0 since the energy bands of the edge states deviate along the $x$ direction in the Brillouin zone. The bottom panels of Figs.~\ref{fig:1} and \ref{fig:2} also show the calculated spin densities, $n_{\uparrow}(r)-n_{\downarrow}(r)$, of the BNC structures. One can see that magnetic ordering appears on the edge of the graphene flakes and that the magnetic moment is strengthened when the electrons around the graphene flakes are localized by the insulating behavior of the BN regions.

We then investigate the reason of the absence of magnetic ordering in models 1(a), 1(b), and 2(a). In studies on metallic magnetism, the Stoner condition provides a natural starting point, $I(E_F) \cdot {\mbox{DOS}}(E_F) > 1$, where $I(E_F)$ is called the Stoner parameter\cite{stoner} and DOS is the density of states at the Fermi level, which are computed in the spin-unpolarized limit. The values of $I(E_F) \cdot {\mbox{DOS}}(E_F)$ are 0.29, 0.58, and 1.23 for models 1(a), 1(b), and 1(c), respectively. In model 1(c), the energy bands of the edge states of the graphene flakes, which play an important in the ferromagnetic ordering, do not vary significantly in the Brillouin zone, and the DOS at the Fermi energy is high. In contrast, the energy bands around the Fermi level are dispersive in models 1(a) and 1(b) owing to the metallic property of the graphene sheet, which results in a spin-unpolarized ground-state electronic structure. On the other hand, the reason for the absence of the magnetic moment in model 2(a) is different from that for Fig.~\ref{fig:1}. Since the BN segment is not sufficiently large in model 2(a), the two energy bands around the Fermi level are separated energetically, and a high DOS at the Fermi level is not observed.

\begin{table}
\caption{Calculated magnetic moments of BNC structures. 1(a), 1(b), 1(c), 2(a), 2(b), 2(c), and 2(d) correspond to structures in Figs.~\ref{fig:1} and \ref{fig:2}.}
\begin{tabular}{ccc}
\hline\hline 
\hspace{1cm} Model \hspace{1cm} & \hspace{3mm} Magnetic moment ($\mu_B$/cell) \hspace{3mm} \\ \hline
1(a) & 0.00   \\
1(b) & 0.00   \\
1(c) & 2.00   \\
2(a) & 0.00   \\
2(b) & 1.39   \\
2(c) & 1.44   \\
2(d) & 1.46   \\
\hline\hline
\end{tabular}
\label{tbl:1}
\end{table}

\subsection{Magnetic moment of G/BNC/G structure}
\label{subsec:gbncg}
\begin{figure}
\includegraphics{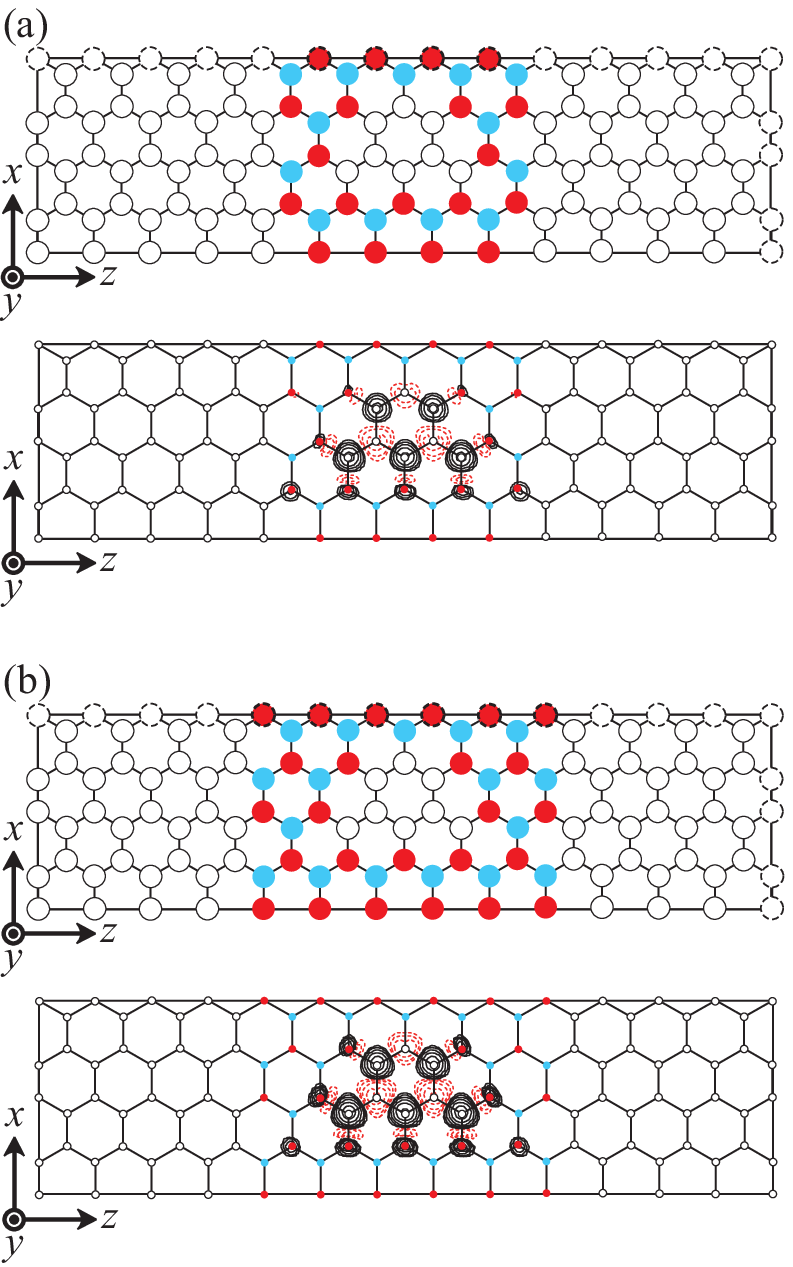}
\caption{(color online) Top view of computational models (top panel) and contour plots showing difference between up-spin and down-spin charge density distributions (bottom panel). The symbols have the same meanings as those in Fig.~\ref{fig:1}.}
\label{fig:3}
\end{figure}

Graphene is one of the best candidates for buffer layers between BNC structures and electrodes because of its metallic characteristics and long-lived spin coherence. We explore the magnetic moment of BNC structures connected to graphene (G/BNC/G structures). Since the transition from ferromagnetic to nonmagnetic states occurs when fewer than 0.05 electrons are doped per BNC structure and the existence of an optimum width for the BN region has been given,\cite{okada} it is of interest whether ferromagnetic states can still be observed in G/BNC/G structures. When primary importance is attached to the magnetic moment of BNC structures, a wider BN region is necessary to obtain a large magnetic moment. However, in terms of the transmission of electrons, a wider BN region acting as an insulator could be an obstacle to obtaining larger conductance. The computational models used to investigate the magnetic moment are outlined in Fig.~\ref{fig:3}. The integration over the Brillouin zone for the $x$ direction is carried out by equidistant sampling of the four $k$-point grid. The calculated total magnetic moments are found to be $1.51$ and 1.62 $\mu_B$/cell for models 3(a) and 3(b), respectively. The magnetic behavior is illustrated by plotting contours of the difference in the charge density distributions on the plane in Fig.~\ref{fig:3}. Note that the spin polarization accumulates in the BNC regions and that the graphene structures as the electrodes do not exhibit magnetic ordering.

\subsection{Transport properties of G/BNC/G structure}
\label{subsec:tra}
It is important to evaluate quantitative spin transmissions toward the application of spin-filter materials. Based on the results in the previous subsection, the spin-transport properties of models 3(a) and 3(b) are investigated. We establish a computational model for the transport calculations, where the G/BNC/G structures are sandwiched between electrodes. Hereafter, we refer to the G/BNC/G structures in Figs.~\ref{fig:3}(a) and \ref{fig:3}(b) connected to {\it semi-infinite} graphene electrodes as models A and B, respectively. The scattering wave functions of the electrons propagating from the left electrode are determined to include the rest of the semi-infinite electrodes by solving the following simultaneous equations for each incident wave function $\Phi^{in}_L(z_k)$:
\begin{equation}
\left[E-\hat{H}_T-\hat{H}_\Sigma\right]\left[
\begin{array}{c}
\Psi(z_{0})   \\
\Psi(z_{1})   \\
\vdots        \\
\Psi(z_{N_z+1}) \\
\end{array}
\right]
=
\left[
\begin{array}{c}
B_z^{\dagger}\Phi^{in}_L(z_{-1})-\Sigma^r_L(z_0)\Phi^{in}_L(z_0) \\ 
    0         \\
\vdots        \\
    0         \\
\end{array}
\right].
\label{eqn07}
\end{equation}
Here, $\hat{H}_T$ is the truncated part of the Hamiltonian of the scattering region and $E$ is the energy of the incident electrons. $\Psi(z_i)$ is the set of values of the scattering wave function on the $x$-$y$ plane at $z=z_i$ and it satisfies the scattering boundary condition, i.e.,
\begin{eqnarray}
\Psi(z_k)=\left\{
\begin{array}{l}
\displaystyle{\Phi^{in}_L(z_k) + \sum_{i=1}^{N_xN_y} r_i \Phi^{ref}_i(z_k)} \hspace{4mm} (k \le 0) \\
\displaystyle{\sum_{i=1}^{N_xN_y} t_i \Phi^{tra}_i(z_k)} \hspace{4mm} (k \ge N_z+1) \hspace{8mm}
\end{array}
\right.,
\label{eqn06}
\end{eqnarray}
where $r_i$ ($t_i$) is the reflection (transmission) coefficient, $\Phi^{ref}_i(z_k)$ $\left(\Phi^{tra}_i(z_k)\right)$ is the generalized Bloch state in the semi-infinite electrodes for reflected (transmitted) electrons, and $N_x$, $N_y$, and $N_z$ are the numbers of grid points in the $x$, $y$, and $z$ directions, respectively. In addition, $\hat{H}_\Sigma$ is a zero matrix except for the first and the last elements, which are the retarded self-energy matrices for the left and right electrodes, $\Sigma^r_L(z_0)$ and $\Sigma^r_R(z_{N_z+1})$, respectively. Further details can be found in Refs.~\onlinecite{obm} and \onlinecite{kong}. The retarded self-energy matrices for the graphene electrodes are employed, and eight graphene buffer layers are inserted between the BNC structure and electrodes. A grid spacing of 0.21 \AA \hspace{1mm} is employed. The scattering wave functions from the right electrode are treated in the same way. We first calculate $\hat{H}_T$ using periodic boundary conditions and then compute the scattering wave functions obtained non-self-consistently. It has been reported that this procedure is just as accurate as fully self-consistent calculations in the linear response regime but significantly more efficient than being self-consistent on a scattering basis.\cite{kong2} The Brillouin zone is sampled with the four $k$-point grid to set up $\hat{H}_T$. The conductance per unit cell under zero temperature and zero bias is described by the Landauer-B\"uttiker formula,\cite{buttiker}
\begin{equation}
G(E)= G_0\int d\vecvar{k}_{||}\frac{A}{(2\pi)^2}\text{Tr}(\textbf{T}^{\dag}\textbf{T}),
\end{equation}
where $\textbf{T}$ is a transmission-coefficient matrix, $A$ is the area of the unit cell in the $x$ and $y$ directions, and G$_0=2e^2/h$ with $e$ and $h$ being the electron charge and Planck's constant, respectively.

\begin{figure}
\includegraphics{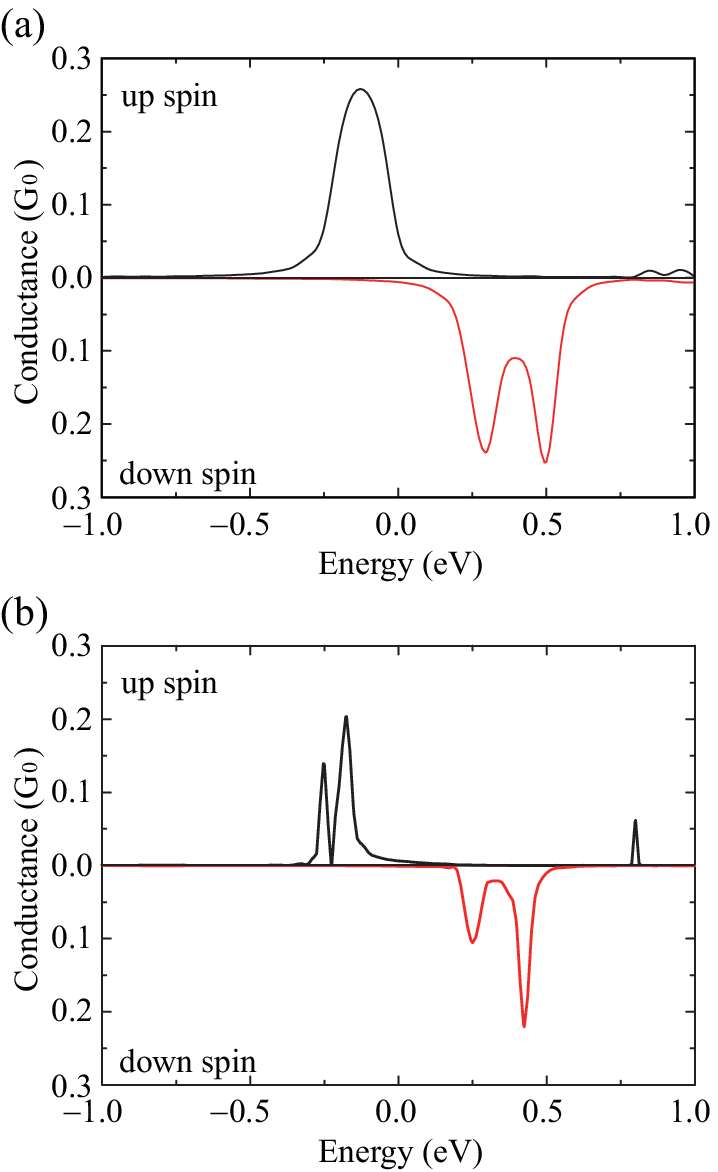}
\caption{(color online) Conductance as a function of energy of incident electrons. Zero energy is chosen to be at the Fermi level.}
\label{fig:4}
\end{figure}

\begin{figure*}
\includegraphics{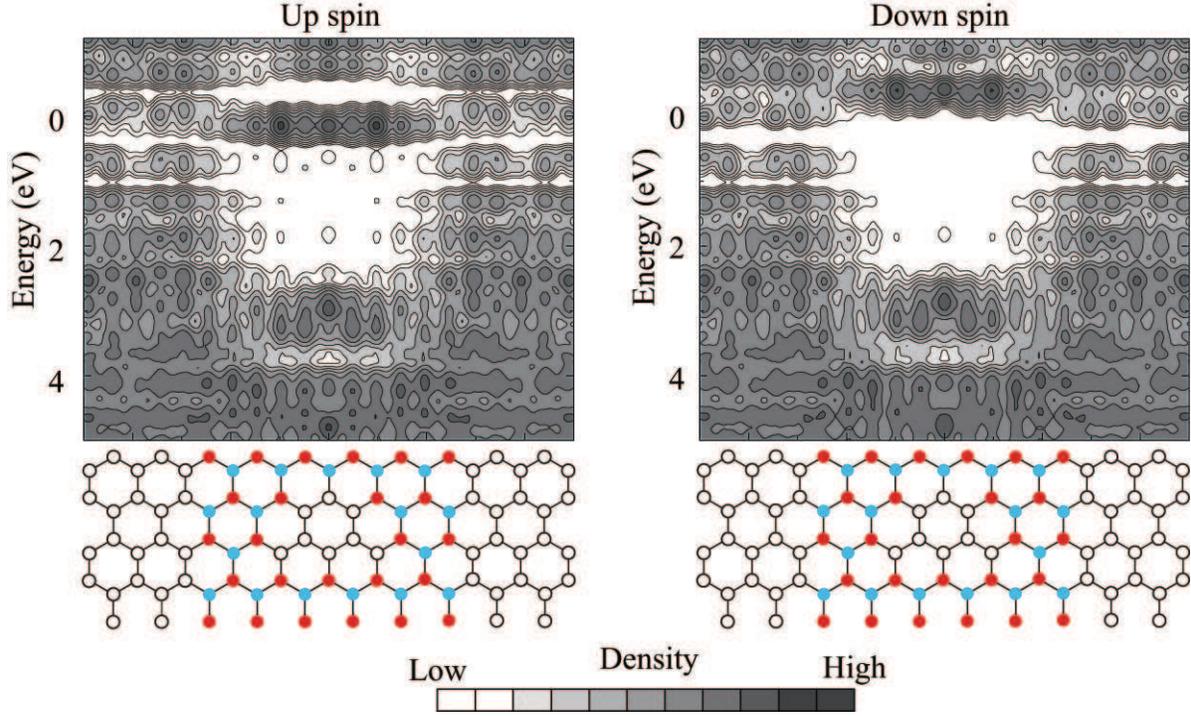}
\caption{(color online) Distributions of LDOS of G/BNC/G structure integrated on plane parallel to interface as functions of relative energy from Fermi level. Zero energy is chosen to be at the Fermi level. Each contour represents twice or half the density of the adjacent contour lines, and the lowest contour is 6.78 $\times 10^{-5}$ {\it e}/eV/\AA. The atomic configurations are given as a visual guide below the graph, and the symbols have the same meanings as those in Fig.~\ref{fig:1}.}
\label{fig:5}
\end{figure*}

\begin{figure}
\includegraphics{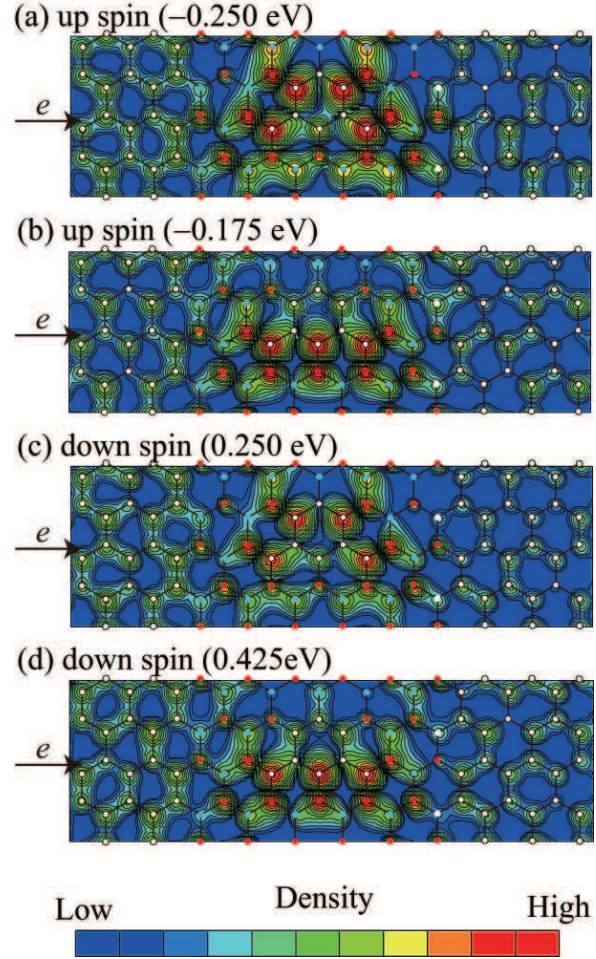}
\caption{(color) Charge density distributions of scattering waves for incident electrons emitted from left electrode on plane parallel to BNC structure of model B. Each contour represents twice or half the density of the adjacent contour lines, and the lowest contour is 1.96 $\times 10^{-3}$ {\it e}/eV/\AA$^3$. The symbols have the same meanings as those in Fig.~\ref{fig:1}. Energies in parentheses are relative to the Fermi level.}
\label{fig:6}
\end{figure}

\begin{figure}
\includegraphics{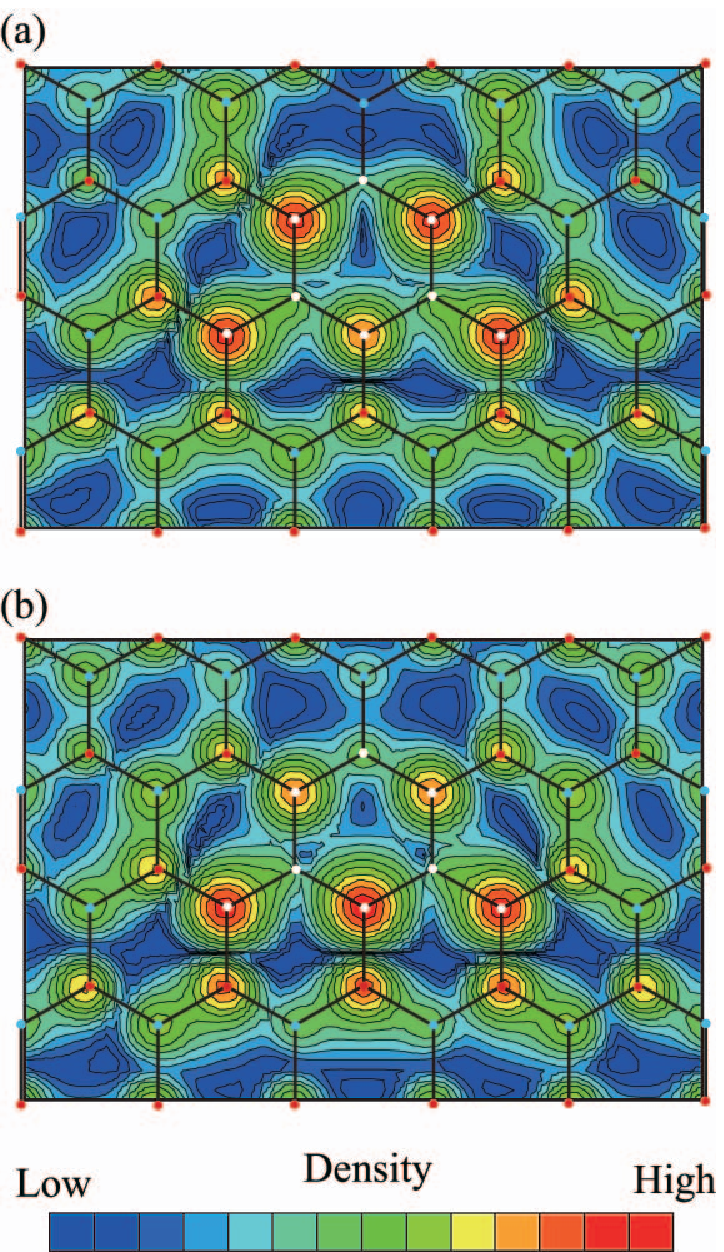}
\caption{(color) Charge density distributions of (a) lower edge state [indicated by $\alpha$] and (b) higher edge state [indicated by $\beta$] around the Fermi level for BNC structure in Fig.~\ref{fig:2}(c). Each contour represents twice or half the density of the adjacent contour lines, and the lowest contour is 1.29 $\times 10^{-5}$ {\it e}/\AA$^3$.}
\label{fig:7}
\end{figure}

Figure~\ref{fig:4} plots the conductance of G/BNC/G structures as a function of the energy of the incident electrons. Although BN sheets are insulators, considerable electron transmission through BNC structures can be observed. The magnitude of conductance for model B is smaller than that for model A because of the insulating behavior of the long BN regions. In addition, the conductance spectrum of model B contains sharp peaks attributed to the resonant tunneling. There is a significant spin-polarized electron current observed that can be associated with the BNC structures. Here, we define the parameter $P (E) = [\sigma_{\uparrow}(E) - \sigma_{\downarrow}(E)]/[\sigma_{\uparrow}(E) + \sigma_{\downarrow}(E)]$ to characterize the spin polarization of electron current, where the conductance of spin $s(=\uparrow,\downarrow)$ is donated by $\sigma_{s} (E)$. The spin-polarization ratio is found to be $\sim$ 1, which is comparable to that obtained with ferromagnetic tunnel junctions using a transition metal.\cite{kokado} The peaks of the conductance of the up-spin and down-spin channels are split by $\sim$ 0.5 eV. The differences between the positions of the peaks for the up-spin and down-spin spectra are similar for models A and B because the splitting of the bands for the up-spin and down-spin edge states is almost equal in Figs.~\ref{fig:3}(c) and \ref{fig:3}(d). Moreover, the energy ranges of $P(E)=1$ and those of $P(E)=-1$ are completely separated. Although graphene is a gapless semiconductor, the conductance of the up spin (down spin) is negligibly small except for in the energy ranges of $P(E)=1$ ($P(E)=1$) because of the finite size effect of the triangular flakes surrounded by the BN sheets. This means that we can control the spin-polarized current by tuning the gate bias.

Figure~\ref{fig:5} shows the local density of states (LDOS) of model B, which is plotted by integrating them along the $x$-$y$ plane, $\rho(z,E)=\int |\psi(\vecvar{r},E)|^2 d\vecvar{r}_{||}$, where $\vecvar{r}=(x,y,z)$, $\psi$ is the wave function, and $E$ is the energy of the states. The states of up-spin electrons are shifted to lower energies and those of down-spin electrons are shifted to higher energies in the BNC structure, although the spin polarization is negligibly small in the LDOS of the graphene region. Thus, the conductance spectrum of the BNC structure is not symmetric around the Fermi level, whereas that for graphene is symmetric. Moreover, the difference in energy between these states of up-spin and down-spin electrons is $\sim$ 0.5 eV. We can see good correspondence between the conductance and LDOS, which implies that the electronic structure of the BNC structure contributes to the spin-polarized electron current.

To investigate the origin of the two significant peaks in the conductance spectrum in Fig.~\ref{fig:4}(b), we show in Fig.~\ref{fig:6} the charge density distribution of scattering waves for model B, in which the distribution of the waves for incident electrons from the left electrode is plotted. In Fig.~\ref{fig:7}, the charge density distribution of the two edge states around the Fermi level in model 2(c) is also depicted. The charge density distribution of the energetically lower edge state [indicated by $\alpha$ in Fig.~\ref{fig:2}(c)] is aligned along a slope of the triangular graphene flake and that of the higher edge state [indicated by $\beta$] is parallel to the base. As can be seen in Fig.~\ref{fig:6}, the lower edge state contributes to electron transport at low energies and the higher edge state contributes to electron transport at high energies in both spin channels.

Such a spintronic device can be compared to another theoretical proposal that achieves a spin-filter material. Compared with the graphene antidot system,\cite{antidot} this BNC structure has a wider energy window where the current is fully polarized. The inserted BN segment between the graphene flakes and electrodes insulates the edge states so that the edge states accumulate in the graphene flakes, which results in a higher magnetic moment for the BNC structure. Our results indicate that the BNC structure is one of the most promising candidates for electronic control over spin transport.

\section{Summary}
\label{sec:summ}
We investigated the electronic structures and transport properties of triangular graphene flakes surrounded by BN sheets using first-principles calculations. We found that the magnetic moment of the graphene flakes increases as the flakes become small and as they are surrounded by a large BN region. When the BNC structure is connected to graphene electrodes, the spin polarization of the charge density distribution accumulates at the edges of the flakes and no spin polarizations is observed in the graphene electrodes. First-principles transport calculation revealed that electron transport through the BNC structure is fully polarized in a wide energy range around the Fermi level. These results should stimulate interest in spin-transport devices using carbon-based materials and bottom-up technology.

\section*{Acknowledgements}
The authors would like to thank Kikuji Hirose and Yoshitada Morikawa of Osaka University for fruitful discussion. This research was partially supported by Strategic Japanese-German Cooperative Program from Japan Science and Technology Agency and Deutsche Forschungsgemeinschaft, by a Grant-in-Aid for Scientific Research on Innovative Areas (Grant No. 22104007) from the Ministry of Education, Culture, Sports, Science and Technology, Japan. The numerical calculation was carried out using the computer facilities of the Institute for Solid State Physics at the University of Tokyo, the Research Center for Computational Science at the National Institute of Natural Science, Center for Computational Sciences at University of Tsukuba, and the Information Synergy Center at Tohoku University.

\end{document}